\documentclass[12pt,a4paper]{article}
\pdfoutput=1
\usepackage{color}
\usepackage{amssymb,amsmath,bm,bbold}
\usepackage{epsf}
\usepackage{epsfig}
\usepackage{relsize}
\usepackage[dvipsnames]{xcolor}
\usepackage[linktoc=page,bookmarks=false,colorlinks=false,linkbordercolor=RoyalBlue,citebordercolor=ForestGreen,urlbordercolor=CornflowerBlue]{hyperref}
\usepackage{latexsym,mathrsfs,dsfont}
\usepackage[normalem]{ulem} 
\usepackage[compress]{cite}
\usepackage{graphicx}
\usepackage{url}
\usepackage{booktabs}
\usepackage{float}
\usepackage{multirow}
\usepackage{changepage}
\usepackage[hypcap]{caption, subcaption}

\usepackage[titles]{tocloft}

\usepackage{tabularx,colortbl}

\usepackage{fancyhdr}
\advance \headheight by 3.0truept       

\allowdisplaybreaks[1]

\definecolor{red}{cmyk}{0,1,1,0.4}
\definecolor{darkgreen}{rgb}{0.0,0.6,0.0}
\definecolor{cDarkGrey}{RGB}{91,91,91}
\definecolor{cGrey}{RGB}{245,243,238}
\definecolor{cBlue}{RGB}{0,110,191}
\definecolor{cLightBlue}{RGB}{214,237,252}
\definecolor{cRed}{RGB}{196,0,100}
\definecolor{cLightRed}{RGB}{254,222,237}
\definecolor{cGreen}{RGB}{0,166,80}
\definecolor{cLightGreen}{RGB}{254,222,237}
\definecolor{cOrange}{RGB}{221,74,44}
\definecolor{cLightOrange}{RGB}{255,215,210}
\definecolor{cPurple}{RGB}{93,35,125}
\definecolor{cLightPurple}{RGB}{241,230,252}
\definecolor{cYellow}{RGB}{252,191,10}
\definecolor{cISSRBlue}{RGB}{0,111,174}
\definecolor{cISSRGrey}{RGB}{167,169,172}

\newcommand{\beq}{\begin{equation}}
\newcommand{\eeq}{\end{equation}}
\newcommand{\be}{\begin{equation}}
\newcommand{\ee}{\end{equation}}
\newcommand{\bi}{\begin{itemize}}
\newcommand{\ei}{\end{itemize}}
\newcommand{\ba}{\begin{array}}
\newcommand{\ea}{\end{array}}
\newcommand{\beqa}{\begin{eqnarray}}
\newcommand{\eeqa}{\end{eqnarray}}
\newcommand{\bea}{\begin{eqnarray}}
\newcommand{\eea}{\end{eqnarray}}
\newcommand{\beqn}{\begin{eqnarray}}
\newcommand{\eeqn}{\end{eqnarray}}

\newcounter{TODO}

\newcommand{\ord}{{\cal O}}

\newcommand{\GeV}{\,\text{GeV}}

\newcommand{\vcb}{|V_{cb}|}
\newcommand{\vtd}{|V_{td}|}
\newcommand{\vub}{|V_{ub}|}
\newcommand{\vts}{|V_{ts}|}
\newcommand{\vus}{|V_{us}|}
\newcommand{\vud}{|V_{ud}|}
\newcommand{\vcd}{|V_{cd}|}

\newcommand{\epe}{\varepsilon'/\varepsilon}

\def\kpn{K^+\rightarrow\pi^+\nu\bar\nu}
\def\klpn{K_{L}\rightarrow\pi^0\nu\bar\nu}

\newcommand{\hatV}{{\hat V}}

\begin{document}

\begin{flushleft}
\end{flushleft}

\vspace{-14mm}
\begin{flushright}
  AJB-23-4
\end{flushright}

\medskip

\begin{center}
{\large\bf\boldmath
  Natural Suppression of FCNCs at the One-Loop Level\\ in  $Z^\prime$ Models with Implications for    K, D and B Decays  
}
\\[1.0cm]
{\bf
    Andrzej~J.~Buras
}\\[0.3cm]

{\small
TUM Institute for Advanced Study,
    Lichtenbergstr. 2a, D-85748 Garching, Germany \\[0.2cm]
Physik Department, TU M\"unchen, James-Franck-Stra{\ss}e, D-85748 Garching, Germany
}
\end{center}

\vskip 0.5cm

\begin{abstract}
  \noindent
    We analyse $Z^\prime$ contributions to FCNC processes at the one-loop
    level. In analogy to the CKM matrix we introduce two  $3\times3$ {\em unitary}     matrices $\hat\Delta_d(Z^\prime)$ and $\hat\Delta_u(Z^\prime)$ which
    are also {\em hermitian}. They
    govern the flavour interactions mediated by $Z^\prime$ between   down-quarks and up-quarks,  respectively, with
    $\hat\Delta_d(Z^\prime)=\hat\Delta_u(Z^\prime)\equiv \hat\Delta_L(Z^\prime)$
    for left-handed currents due to the unbroken $\text{SU(2)}_L$ gauge symmetry. This assures the suppression of these contributions to all  $Z^\prime$ mediated  FCNC processes at the one-loop level. 
    As, in contrast to the GIM mechanism, one-loop $Z^\prime$ contributions to flavour observables in $K$ and $B_{s,d}$ systems  are   governed by
    down-quark masses, they are ${\cal O}(m^2_b/M^2_{Z^\prime})$ and
   negligible. With the  ${\cal O}(m^2_t/M^2_{Z^\prime})$ suppression
  they are likely negligible  also in the $D$ system. We present an explicit parametrization  of $\hat\Delta_L(Z^\prime)$ in terms of {\em two} mixing angles and {\em two} complex phases that distinguishes it profoundly
  from the CKM matrix.
  This framework can be generalized to purely
  leptonic decays with matrices analogous to the PMNS matrix but profoundly
  different from it. Interestingly, the breakdown of flavour universality between the first two generations and the third one, both for quark and lepton couplings to $Z^\prime$, is identified 
  as a consequence of $\hat\Delta_L(Z^\prime)$
  being  hermitian. The importance of the unitarity for both
  $\hat\Delta_L(Z^\prime)$ and the CKM matrix  in the light of the Cabibbo anomaly is emphasized.
 \end{abstract}

\thispagestyle{empty}
\newpage
\setcounter{page}{1}

%
%
%
\section{Introduction}
As  demonstrated in \cite{Buras:2022wpw}  the $\Delta F=2$
observables  can be consistently
  described within the Standard Model (SM) without
any need for new physics (NP) contributions. In turn 
this allowed to determine precisely the CKM matrix on
the basis of $\Delta F=2$  observables alone \cite{Buras:2022wpw,Buras:2021nns} without the need to face the tensions in
$\vcb$ and $\vub$ determinations from inclusive end exclusive tree-level decays
\cite{Bordone:2021oof,FlavourLatticeAveragingGroupFLAG:2021npn}.
Moreover, as pointed out in \cite{Buras:2022qip}, this also
avoids, under the assumption of negligible NP contributions to these observables, the impact of NP on the values of CKM parameters present likely in global fits. Simultaneously it   provides SM predictions for numerous rare $K$ and $B$  branching ratios that are most accurate to date. In this manner the size of the experimentally observed  deviations from SM predictions (the pulls) can be more reliably  estimated than in global fits.

Popular NP scenarios which are used for the explanation
of various anomalies observed in $\Delta F=1$  observables
are $Z^\prime$ scenarios that usually are studied including only tree-level contributions. 
A natural question then arises what happens at the one-loop level in these
scenarios  under the assumption that in the one-loop diagrams only
$Z^\prime$ gauge bosons and quarks are exchanged\footnote{In a concrete UV completion other particles could contribute as well. But in this paper we assume that their contributions are negligible.}. To my knowledge there is no analysis within $Z^\prime$ models in the literature that addressed this question so far,  because it was always assumed that the tree-level contributions in these models are much more important. However, in a quantum field theory one has
to check what happens at least at the one-loop level. The pioneering
analyses in this context in the case of charged currents mediated by $W^\pm$
bosons have been performed roughly 50 years ago by
Glashow, Iliopoulos and Maiani \cite{Glashow:1970gm} and Gaillard and Lee \cite{Gaillard:1974hs}, who stressed the crucial role of the unitarity of
the Cabibbo-Kobayashi-Maskawa matrix (CKM) \cite{Cabibbo:1963yz,Kobayashi:1973fv} in the suppression of FCNC processes mediated by $W^\pm$ and quarks at the
one-loop level.

The main goal of the present paper is to perform a similar analysis 
for $Z^\prime$ interactions which will lead us to a new matrix, analogous
to the CKM matrix, but this time governing  flavour changing
interactions mediated by neutral $Z^\prime$ gauge bosons instead of
charged current interactions mediated by charged $W^\pm$ bosons. This
will allow us to demonstrate that $Z^\prime$ contributions to FCNC processes at the one-loop level are indeed negligible. However, as we will see below,
this step brings other advantages for $Z^\prime$ models.

Indeed, one of the weak points of most $Z^\prime$ scenarios in the literature
is the absence of correlations between $B_s$, $B_d$, $K$ and $D$ systems.
The prominent exceptions are 331 models based on the gauge group $\text{SU(3})_C\times \text{SU(3)}_L\times \text{U(1)}_X$   \cite{Pisano:1991ee,Frampton:1992wt} that have been
analysed in numerous papers 
with the most recent ones in \cite{Pleitez:2021abk,CarcamoHernandez:2022fvl,Buras:2023ldz}. In these models, as stressed in particular in \cite{Buras:2012dp},
once the new parameters are determined through quark mixing observables in
$B_s$ and $B_d$ systems, no new free parameters are required to describe
the kaon system and subsequently charm system \cite{Colangelo:2021myn,Buras:2021rdg}.

But in all  these analyses only tree-level contributions have been analysed.
Here we would like to make one step further and putting 331 models aside,
generalize the common $Z^\prime$ scenarios beyond the tree-level.
Specifically, we want to construct a framework, in analogy to the CKM theory
\cite{Cabibbo:1963yz,Kobayashi:1973fv} that deals with charged $W^\pm$ interactions, in which flavour violation is governed  by neutral $Z^\prime$ exchanges. To this end we proceed in the following four steps.

{\bf Step 1}

We introduce two  $3\times3$ unitary 
 matrices $\hat\Delta_d(Z^\prime)$ and $\hat\Delta_u(Z^\prime)$ that in contrast to the CKM matrix are also {\em hermitian}. They
 govern the flavour interactions mediated by $Z^\prime$ between   down-quarks and up-quarks,  respectively. They can be used for both left-handed and right-handed currents but for the left-handed ones $\hat\Delta_d(Z^\prime)=\hat\Delta_u(Z^\prime)$ due to  the unbroken ${\rm SU(2)}_L$ gauge symmetry in the SMEFT.

 {\bf Step 2}

 We demonstrate, in analogy to  the GIM mechanism
  \cite{Glashow:1970gm}, that having these two matrices 
 assures the suppression of $Z^\prime$ contributions to all  FCNC processes at the one-loop level. 
  As, in contrast to the GIM mechanism, $Z^\prime$ contributions to flavour observables in $K$ and $B_{s,d}$ systems  are   governed by down-quark masses
  at the one-loop level, these   contributions being $\ord(m^2_b/M^2_{Z^\prime})$
  are negligible. With the  $\ord(m^2_t/M^2_{Z^\prime})$ suppression
  they are strongly suppressed also in the $D$ system. This confirms
  the expected smallness of such contributions.

  {\bf Step 3}

  Most importantly we find a parametrization of these two matrices
  that is analogous to the standard parametrization of the CKM matrix
  \cite{Chau:1984fp}, but due to the additional property of hermiticity it
 differs profoundly from the  latter parametrization. In particular 
 each of these two matrices depends on   {\em two  mixing angles} and not three and {\em two complex phases} and not one as is the case of the CKM matrix.
 For the case of only left-handed couplings these matrices are equal to each other so  that the number of flavour changing parameters, two mixing angles and two phases is as  in the 331 models.

 {\bf Step 4}

 These three steps can be used also for leptons with matrices analogous to the PMNS matrix but profoundly  different from it.

Our paper is organized as follows. In Section~\ref{loopnew} we  describe   the general structure of one-loop contributions to flavour observables in this  NP scenario, stressing the differences from the GIM mechanism \cite{Glashow:1970gm}. In Section~\ref{Parametrization} we present one possible parametrization of
$\hat\Delta_d(Z^\prime)$ in terms of two mixing angles and 
 two complex phases. This parametrization differs profoundly
from  the standard one of the CKM
matrix. Analogous parametrization can be used for  $\hat\Delta_u(Z^\prime)$ and
also for purely leptonic processes. In the latter case it differs profoundly from the Pontecorvo-Maki-Nakagawa-Sakata (PMNS) matrix \cite{Pontecorvo:1957cp,Maki:1962mu}.
In Section~\ref{Correlations} we discuss briefly the implied flavour patterns in  $K$ and $B$ decay   observables in this NP  scenario, in particular correlations between various branching ratios. In Section~\ref{NLAST} we address the
popular topic of flavour universality violation both in the quark and the lepton sector in our framework. In Section~\ref{2NLAST}, motivated by the numerous
papers on the Cabibbo anomaly, we stress the importance of
the unitarity for both $\hat\Delta_L(Z^\prime)$ and the CKM matrix  that is
crucial for obtaining gauge independent results at the one-loop level.
We conclude in Section~\ref{LAST}.

\boldmath
\section{Natural Suppression of One-Loop $Z^\prime$ Contributions to FCNC Processes }\label{loopnew}
\unboldmath
In the SM the GIM mechanism \cite{Glashow:1970gm} is violated in FCNC processes in $K$ and $B$ systems
at the one-loop level through the virtual contributions of the top-quark in
box and penguin diagrams. In the charm meson system it is still satisfied to
a high degree because only down quarks are exchanged together with $W^\pm$ bosons
in one-loop diagrams. This leads to an additional suppression of these
contributions by factors like $m_i^2/M_W^2$ with $i=d,s,b$.

Here we would like to point out that in the case of $Z^\prime$ models the situation
is reversed. In the  $K$ and $B$ systems only down-quarks are exchanged together with $Z^\prime$ bosons in one-loop diagrams leading to an additional suppression of these contributions by factors like $m_i^2/M^2_{Z^\prime}$ with $i=d,s,b$.
Consequently, the tree-level contributions of $Z^\prime$ gauge bosons
fully dominate the $K$ and $B$ decays.
In the charm system top-quark contributions at the one-loop level
could in principle play some role although this time they are suppressed
by  $m_t^2/M^2_{Z^\prime}$ and likely small.

For the $K$ and $B_{s,d}$ systems we assume that
FCNC processes mediated by a $Z^\prime$ gauge boson are governed by a unitary
and hermitian $3\times 3$ matrix
\begin{align}\label{neutralGIM}
  \hat \Delta_d =\left(\begin{array}{ccc}
							\Delta_{dd}&\Delta_{ds}&\Delta_{db}\\
							\Delta_{sd}&\Delta_{ss}&\Delta_{sb}\\
							\Delta_{bd}&\Delta_{bs}&\Delta_{bb}
							\end{array}\right)\,,\qquad \hat\Delta_d^\dagger \hat \Delta_d=\hat 1,\qquad \Delta_{ij}=(\Delta_{ji})^*\,.
\end{align}
\noindent
Similar for the up-quark system we introduce
\begin{align}\label{neutralGIM2}
  \hat \Delta_u =\left(\begin{array}{ccc}
							\Delta_{uu}&\Delta_{uc}&\Delta_{ut}\\
							\Delta_{cu}&\Delta_{cc}&\Delta_{ct}\\
							\Delta_{tu}&\Delta_{tc}&\Delta_{tt}
							\end{array}\right)\,,\qquad \hat\Delta_u^\dagger \hat \Delta_u=\hat 1,\qquad \Delta_{ij}=(\Delta_{ji})^*\,.
\end{align}

This implies
\be\label{UH}
\hat \Delta_d^2=\hat 1, \qquad \hat \Delta_u^2=\hat 1\,.
\ee

These matrices can be introduced for both left-handed and right-handed $Z^\prime$ couplings but in what follows we will confine our discussion to left-handed
couplings, so that in this case
\be\label{SU2}
\hat\Delta_d(Z^\prime)=\hat\Delta_u(Z^\prime)\equiv\hat\Delta,\qquad ({\rm SU(2)}_L),
\ee
due to  the unbroken ${\rm SU(2)}_L$ gauge symmetry in the SMEFT for
$Z^\prime$ masses above a few TeV.

The matrix $ \hat \Delta$ is analogous to the CKM matrix, but in contrast to the latter
involves only transitions between down-quarks and between up-quarks. Very importantly it is hermitian, which is not the case of the CKM matrix.  In practice, in the vertices involving $Z^\prime$
and two quarks the relevant elements of the matrix $\hat \Delta$ are multiplied by a flavour universal
coupling like $g_2$ in the SM, but setting this coupling to unity simplifies
the formulae below without changing the argument.

In order to demonstrate that indeed in the limit of degenerate down-quark masses
($m_d=m_s=m_b$) the one-loop contributions to all FCNC transitions in $Z^\prime$
models  vanish, it is sufficient
to consider a single fermion line in a loop diagram with two $Z^\prime$
vertices. Considering the $s\to b$ transition, the three contributions from
the down quarks are
\be
s\to d\to b, \qquad s\to s\to b, \qquad s\to b\to b\,,
\ee
implying in the mass degeneracy limit the overall factor in the one-loop NP contribution
to $M_{12}^{bs}$\footnote{In each term in the first sum the last flavour index on the right is the incoming   $s$ and the first one on the left the outgoing $b$.}
\be\label{23}
\boxed{\Delta_{bd}\Delta_{ds}+\Delta_{bs}\Delta_{ss}+\Delta_{bb}\Delta_{bs}=
\Delta_{bd}\Delta_{sd}^*+\Delta_{bs}\Delta_{ss}^*+\Delta_{bb}\Delta_{sb}^*=0.}
\ee
The first relation follows from the hermiticity of $\hat\Delta$ 
in (\ref{neutralGIM}), while the second one from its unitarity.

Similar, for $d\to b$ transitions we have
\be
d\to d\to b, \qquad d\to s\to b, \qquad d\to b\to b\,,
\ee
implying in the mass degenaracy limit the overall factor in   the one-loop NP contribution
to $M_{12}^{bd}$
\be\label{13}
\boxed{\Delta_{bd}\Delta_{dd}+\Delta_{bs}\Delta_{sd}+\Delta_{bb}\Delta_{bd}=
\Delta_{bd}\Delta_{dd}^*+\Delta_{bs}\Delta_{ds}^*+\Delta_{bb}\Delta_{db}^*=0.}
\ee

Finally, for $d\to s$ transitions we have
\be
d\to d\to s, \qquad d\to s\to s, \qquad d\to b\to s\,,
\ee
and the overall factor in  the one-loop NP contribution
to $M_{12}^{sd}$
\be\label{12}
\boxed{\Delta_{sd}\Delta_{dd}+\Delta_{ss}\Delta_{sd}+\Delta_{sb}\Delta_{bd}=
\Delta_{sd}\Delta_{dd}^*+\Delta_{ss}\Delta_{ds}^*+\Delta_{sb}\Delta_{db}^*=0.}
\ee
Beyond this limit in all these cases the leading NP contributions are $\ord(m_b^2/M^2_{Z^\prime})$ and can be safely neglected. As discussed in Section~\ref{2NLAST}, explicit calculations in the unitary gauge confirm this expectation. Analogous relations can be derived for the up-quark system. However, as reemphasized
 in Section~\ref{2NLAST}, the suppression in question
takes place {\em only} due to the unitarity of $\hat\Delta$.

The three relations in (\ref{23}), (\ref{13}) and (\ref{12}) and the corresponding ones for the up-quarks correspond to
the usual six unitarity triangles resulting from the unitarity of the
CKM matrix.  Yet, they differ from the latter in a profound manner:
\begin{itemize}
\item
  They are only three for $K$ and $B$ observables and not six because of the hermiticity property on the top   of the unitarity of  $\hat\Delta$.
\item
  Because of the hermiticity of $\hat\Delta$, the phases of its elements
  drop out   in these relations which will be easily seen using the parametrization of $\hat\Delta$ in (\ref{B1}). Consequently there are no unitarity triangles of the type known from the   CKM-phenomenology.
\end{itemize}

However, these relations put important constraints on $\hat\Delta$ and deserve
a name to be distinguished from the usual unitarity triangle relations.
They could be called {\em Unitarity-Hermiticity-Triangles (UHT)}.

\boldmath
\section{A Parametrization of $\hat\Delta$}\label{Parametrization}
\unboldmath
\subsection{New Mixing Matrix}
We have just seen that the matrix $\hat\Delta$ in (\ref{neutralGIM}) played the crucial role in the suppression of FCNC processes  generated by neutral gauge bosons at the one-loop level. This is analogous to the CKM matrix that in the context of the GIM mechanism allows to suppress such processes generated by $W^\pm$ at the one-loop level. But analogous to the GIM mechanism, that still allows tree-level contributions mediated by $W^\pm$ at tree-level and is governed by the CKM matrix, the matrix $\hat\Delta$ governs FCNC processes mediated by $Z^\prime$ at tree-level. 

For phenomenology it is important to find a useful parametrization for $\hat\Delta$. We will
attempt to do it below being aware of the fact that dependently on the
future status of experimental data our parametrization could be replaced by a
more useful one one day. In fact the original parametrization of the CKM matrix
by Kobayashi and Maskawa \cite{Kobayashi:1973fv} turned out not to be useful
for phenomenology and was replaced in 1984 by a different one \cite{Chau:1984fp}, known these days under the name of {\em the standard  parametrization of the CKM matrix.}

In order to find the parametrization of $\hat\Delta$ we impose the usual
unitarity conditions. But
 due to the hermiticity of $\hat\Delta$, we have in addition
to the usual  three independent unitarity constraints on the moduli of the elements of this matrix, 
\be
|\Delta_{dd}|^2+|\Delta_{ds}|^2+|\Delta_{db}|^2=1,
\ee
\be
|\Delta_{sd}|^2+|\Delta_{ss}|^2+|\Delta_{sb}|^2=1,
\ee
\be
|\Delta_{bd}|^2+|\Delta_{bs}|^2+|\Delta_{bb}|^2=1,
\ee
{\em only} three UHT relations given in (\ref{23}) and (\ref{13})
and (\ref{12}).

Before studying these constraints it is useful to construct first  the $2\times2$
matrix that is analogous to the Cabibbo matrix.  We find
\be\label{22}
\hat\Delta = \begin{pmatrix}
                 c & se^{-i\delta} \\
		se^{i\delta}    & -c
                \end{pmatrix}, 
\ee
with $s=\sin\theta$ and $c=\cos\theta$,
signalling that the $3\times3$ matrix, we are looking for, will be quite
different from the CKM matrix. Not only we have already a new phase in this matrix but also the signs are distributed differently from the Cabibbo matrix.

In searching for a useful parametrization of $\hat\Delta$ I was guided by the
the structure of $Z^\prime$ interactions used in numerous collaborations
with Fulvia De Fazio and Jennifer Girrbach-Noe in the context of the 331 models
\cite{Buras:2012dp,Buras:2013dea,Buras:2014yna,Buras:2015kwd,Buras:2016dxz,Buras:2021rdg}. However, in these models, that contain additional heavy fermions,
some phases could be rotated away from the flavour violating couplings. Moreover, working at
tree-level, no attempt has been made to construct a matrix like $\hat\Delta$.

In our case, in the absence of new quark fields beyond the SM quarks, we are not allowed to perform phase rotations on the quark fields as done in the process of constructing the CKM matrix, because these rotations have already been made in the latter construction. However, imposing hermiticity on $\hat\Delta$ in
addition to unitarity, our matrix depends  on four parameters,
two mixing angles and two complex phases that we denote by 
\be\label{PARB}
\theta_1,\qquad \theta_2,\qquad  \delta_1,\qquad \delta_2,
\ee
and express $\hat\Delta$ through
\be\label{PARB2}
s_1=\sin\theta_1,\qquad c_1=\cos\theta_1,\qquad s_2=\sin\theta_2,\qquad c_2=\cos\theta_2,
\ee
and the phases $\delta_1$ and $\delta_2$. Also  flavour violating couplings
of $Z^\prime$ gauge boson in 331 models depend on two mixing angles and two
complex phases.

The number of parameters is then the same as in the CKM matrix but the presence
of two complex phases instead of one promises larger CP-violating effects
than in the SM. Similar to the SM and MFV models, the paucity of free parameters in $\hat\Delta$ implies correlations between various observables.
But these correlations are different from the MFV-ones, in particular because
of two phases and not only one and because of only two mixing angles and not
three.

The construction of a $3\times3$ matrix that is both unitary and hermitian
is more challenging than for the $2\times2$ one and I did not find any example in the literature. With my experience in 331 models, the structure of phases in this matrix was evident, but finding the coefficients in front of the phase factors required more work\footnote{In this context  invaluable advices  from Herbert Spohn and Peter Weisz should be mentioned.}.
We arrive then at
\be\label{B1}
\boxed{\hat\Delta = \begin{pmatrix}
             1-2 s_1^2s_2^2  & -2 s_1^2 s_2 c_2 e^{-i(\delta_1-\delta_2)}  & -2s_1 s_2 c_1  e^{-i\delta_1} \\ - 2 s_1^2 s_2 c_2 e^{i(\delta_1-\delta_2)}
             & 1-2 s_1^2c^2_2 & -2s_1c_1c_2  e^{-i\delta_2}     \\

-2s_1 s_2 c_1  e^{i\delta_1}	     &   -2s_1c_1c_2        e^{i\delta_2}             & 1-2 c_1^2
                \end{pmatrix},}
\ee
which is the main result of our paper, the construction of an analog of the CKM
matrix for processes  mediated by neutral $Z^\prime$
gauge bosons as opposed to $W^\pm$ in the case of the CKM matrix. The
important virtue of this matrix are correlations between various meson systems
that follow from the unitarity and hermiticity of this matrix and 
implied  suppression of FCNCs at the one-loop level. This is
similar to the unitarity of the CKM matrix which implies various correlations
between decay branching ratios of different mesons as well as
the GIM mechanism.

The three UHT relations  in (\ref{23}) and (\ref{13})
and (\ref{12}) can now be written respectively, as follows
\be
\frac{|\Delta_{bd}||\Delta_{ds}|}{|\Delta_{bs}||\Delta_{ss}|} +
\frac{|\Delta_{bb}}{|\Delta_{ss}|} =1\,,
\ee
\be
\frac{|\Delta_{bs}||\Delta_{sd}|}{|\Delta_{bd}||\Delta_{dd}|} +
\frac{|\Delta_{bb}}{|\Delta_{dd}|} =1\,,
\ee
\be
\frac{|\Delta_{sb}||\Delta_{bd}|}{|\Delta_{ss}||\Delta_{sd}|} -
\frac{|\Delta_{dd}}{|\Delta_{ss}|} =1\,.
\ee

One can easily check that indeed the condition $\hat\Delta^2=\hat 1$ is satisfied. One can also check that the phases $\delta_1$ and $\delta_2$ drop out from
 the relations  in (\ref{23}) and (\ref{13})
 and (\ref{12}), so there are no unitarity triangles of the CKM-type in this framework. But as these relations  play a crucial role in the suppression of $Z^\prime$ one-loop
 contributions to FCNC processes, we termed them  UHT and expressed them in
 a different form above.
Note that after the phases dropped out  these relations are satisfied by only two mixing angles $\theta_1$ and $\theta_2$.

An analogous matrix can be constructed for the up-quark sector with parameters
that can differ from the ones in the down-quark sector in the case
of right-handed currents but the same for left-handed currents due
to ${\rm SU(2)}_L$ gauge symmetry in the SMEFT.
In the latter case, this implies correlations between $(K,B)$ and $D$ systems,
which however differ from the ones in 331 models
\cite{Colangelo:2021myn,Buras:2021rdg}, simply because the couplings
given in terms of $s_1$ and $s_2$ differ in the 331 models from the ones in
(\ref{B1}). This originates in the fact that there are additional heavy
fermions in the latter models.

Comparing with the standard parametrization of the unitary CKM matrix \cite{Chau:1984fp}
\be\label{equ:VCKMWinkel}
V_\text{CKM} = \begin{pmatrix}
                c_{12} c_{13} & s_{12} c_{13} & s_{13} e^{-i\delta} \\
		-s_{12}c_{23}-c_{12} s_{23} s_{13} e^{i\delta} & c_{12} c_{23} - s_{12} s_{23} s_{13} e^{i\delta} & s_{23} 
c_{13}\\
		s_{12} s_{23}- c_{12} c_{23} s_{13} e^{i\delta} & -c_{12} s_{23} - s_{12} c_{23} s_{13} e^{i\delta} & c_{23} 
c_{13}
                \end{pmatrix},
\ee
we note  the impact of the hermiticity of $\hat \Delta$. In particular, in contrast to the CKM matrix
\begin{itemize}
\item
  All non-diagonal entries in $\hat\Delta$ are complex.
\item
  All diagonal entries are real but the universality of diagonal couplings in
  the presence of small mixing angles, that is present in the case of the CKM matrix,  is
  broken in the case of $\hat \Delta$ as seen in (\ref{B1A}) below and discussed in Section~\ref{NLAST}.
\item
  $|\Delta_{ij}|=|\Delta_{ji}|$, which is not the case for the CKM matrix as $\vub\not=\vtd$
  and $\vcb\not=\vts$.
\end{itemize}

As $s_1$ and $s_2$ are expected to be  small, it is
useful to approximate $c_1$ and $c_2$ by unity. In doing this  we consider
mixing angles in the first quadrant but then $\delta_i \in [ 0\,,2\pi]$.
We find then
\be\label{B1A}
\hat\Delta \approx \begin{pmatrix}
             1  & -2 s_1^2 s_2  e^{-i(\delta_1-\delta_2)}  & -2s_1 s_2   e^{-i\delta_1} \\ - 2 s_1^2 s_2  e^{i(\delta_1-\delta_2)}
             & 1 & -2s_1  e^{-i\delta_2}     \\
-2s_1 s_2   e^{i\delta_1}	     &   -2s_1  e^{i\delta_2}             & -1
                \end{pmatrix}.
\ee
This implies
\be
s_1\approx\left|\frac{\Delta_{ds}}{\Delta_{db}}\right|,\qquad s_2\approx\left|\frac{\Delta_{db}}{\Delta_{sb}}\right|\,.
\ee

\subsection{Determination of the Parameters}
Let us first count the number of parameters we have to determine from
$\Delta F=1$ and $\Delta F=2$ processes involving quarks. We have
\begin{itemize}
\item
  4 CKM parameters.
\item
  4 parameters of $\hat\Delta_d(Z^\prime)=\hat\Delta_u(Z^\prime)$ for
  {\em left-handed couplings}. It includes also flavour diagonal left-handed quark couplings.
\item
  We need also diagonal {\em right-handed} couplings in order to generate
  NP contributions to $\epe$ through electroweak penguins, the most efficient method to affect $\epe$ \cite{Buras:2015jaq,Aebischer:2020mkv}.  They are universal   within up-quark couplings and universal within down quark coupling and related
  through
  \be
  \Delta_R^{dd}(Z^\prime)=-\frac{1}{2}\Delta_R^{uu}(Z^\prime),
      \ee
      which brings an additional  parameter. As discussed in
      Section~\ref{NLAST}, this
      universality can be arranged in $Z^\prime$ models with specific values of
      $s_i$.
    \item
      Lepton sector brings additional parameters. Their number  depends on the assumption of the flavour universality for all generations or only for 
      the first two. It also depends on the presence of 
right-handed neutrinos. In any case parameters, analogous to the PMNS ones, enter the
    game and we will comment on them below.
  \item
    Finally we have the mass $M_{Z^\prime}$.
\end{itemize}

Definitely the number of parameters is above 10.
These looks at first sight like many, but with the developed codes like
Flavio \cite{Straub:2018kue} and HEPfit  \cite{DeBlas:2019ehy}, that
include a large  number of observables,  their determination should be doable.
Even more important would be the construction of concrete UV completions
which would allow to reduce the number of parameters and  make more concrete predictions for FCNC observables than we can do in the simple framework presented here.

\subsection{Lepton Flavour Violation}
These ideas can be extended to lepton flavour violating processes. In contrast
to the PMNS matrix that contains two additional Majorana phases relative to the CKM
matrix, the structure of the $Z^\prime$  matrix is the same as the one
of $\hat\Delta_{d}$ and $\hat\Delta_{u}$. The reason is that in constructing
$\hat\Delta$ no phase rotations on the fields are allowed because they
have been already made when constructing the PMNS matrix. Thus only
the unitarity and hermiticity properties matter. For instance in the case
of charged leptons the matrix

\begin{align}\label{neutralGIM3}
  \hat \Delta_\ell =\left(\begin{array}{ccc}
							\Delta_{ee}&\Delta_{e\mu}&\Delta_{e\tau}\\
							\Delta_{\mu e}&\Delta_{\mu\mu}&\Delta_{\mu\tau}\\
							\Delta_{\tau e}&\Delta_{\tau\mu}&\Delta_{\tau\tau}
							\end{array}\right)\,,\qquad \hat\Delta_\ell^\dagger \hat \Delta_\ell=\hat 1,\qquad \Delta_{ij}=(\Delta_{ji})^*\,.
\end{align}
is involved. Its parametrization is like the one in (\ref{B1}), only mixing angles and phases are different.
Of interest is
also the breakdown of the universality of $Z^\prime$ couplings to $(e,\mu)$ and $\tau$ couplings analogous to the one in charged currents represented 
through $R(D)$ and $R(D^*)$ ratios. We will discuss this
briefly in Section~\ref{NLAST}.

\section{Correlations between Branching Ratios}\label{Correlations}
The structure of the $\hat\Delta (Z^\prime)$ matrix implies correlations
separately both within $K$, $B_d$ and $B_s$ systems and also between different
meson systems. This is analogous to the correlations between various observables
within the SM,  MFV models and 331 models \cite{Buras:2012dp,Buras:2023ldz}.
\begin{itemize}
\item
  The correlations within the $K$ system will be governed by the element
  $\Delta_{ds}$. For $\delta_1\not=\delta_2$ correlations between 
$\kpn$, $\klpn$, $K_S\to\mu^+\mu^-$, $K_L\to\pi^0\ell^+\ell^-$ and $\epe$
  will exist, similar to those consider recently in  \cite{Aebischer:2023mbz}.
  In the context of  our parametrization of $\hat\Delta$, the choice of phases
  made by these authors corresponds to setting
  $\delta_2-\delta_1=90^\circ$. This allows  to avoid NP contributions to $\varepsilon_K$ and reduces the number of free parameters to three, one phase and
  two mixing angles.
\item
  The correlations within the $B_d$ system are described by the element $\Delta_{db}$ and the ones in the $B_s$-system by $\Delta_{sb}$.
\item
  Similar to 331 models there are correlations between $K$, $B_d$ and $B_s$
  observables which is best seen using the approximation for $\hat\Delta$
  in (\ref{B1A}) which implies
  \be\label{FIT}
  \boxed{\Delta_{db}\Delta_{bs}=-2 \Delta_{ds}.}
  \ee
  \end{itemize}
These are just few examples which we plan to study in more detail elsewhere.

It should finally be noted that due to the relation (\ref{FIT}) the full matrix can be found by determining $\Delta_{db}$ and $\Delta_{sb}$ from  
$B_d$ and $B_s$ decays, respectively.  In turn  Kaon observables can be predicted. This is also possible in 331 models \cite{Buras:2012dp,Buras:2013dea,Buras:2014yna,Buras:2015kwd,Buras:2016dxz,Buras:2021rdg,Buras:2023ldz} but
the structure of the correlations is different due to  different structure of the couplings.  We will report on them  in a future numerical analysis.

\section{Quark and Lepton Flavour Universality Violation}\label{NLAST}
The approximation in (\ref{B1A}) exibits very clearly the universality
of the diagonal couplings of $Z^\prime$ to down-quark and strange-quark  for small $s_i$. However,  this  universality is broken between the couplings of the latter quarks and the $b$-quark.  Similar pattern is observed in the up-quark sector. This different behaviour of the third generation is known from 331 models.

In the $Z^\prime$ scenarios discussed here,
the only values of $s_i$ and $c_i$ for which the universality of the $Z^\prime$ couplings for three generations is achieved are
\be
s^2_1=\frac{2}{3}, \qquad c^2_1=\frac{1}{3},\qquad s^2_2=c^2_2=\frac{1}{2}
\ee
in which case the matrix $\hat\Delta$ depends only on the phases $\delta_1$ and $\delta_2$ and is given by
\be\label{B1U}
\hat\Delta = \begin{pmatrix}
             \frac{1}{3}  & -\frac{2}{3} e^{-i(\delta_1-\delta_2)}  & -\frac{2}{3}  e^{-i\delta_1} \\  -\frac{2}{3}e^{i(\delta_1-\delta_2)}
             & \frac{1}{3}  & -\frac{2}{3} e^{-i\delta_2}     \\

-\frac{2}{3} e^{i\delta_1}	     &   -\frac{2}{3}       e^{i\delta_2}             & \frac{1}{3} 
                \end{pmatrix}.
\ee
It will be interesting to investigate these results phenomenologically.

\boldmath
\section{Gauge Invariance and the Unitarity of $\hat\Delta(Z^\prime)$ and $V_\text{CKM}$ Matrices}\label{2NLAST}
\unboldmath
There was recently a large activity in connection with the so-called Cabibbo
anomaly (CAA) which is nicely reviewed in  \cite{Crivellin:2022ctt}.
 There exist tensions between
different determinations of the elements $\vus$ and $\vud$ in the CKM matrix from different decays. For instance, the determination of $\vud$ from superallowed beta decays and of $\vus$ from kaon decays imply a violation of the first row
unitarity in the CKM matrix. There exist also tensions between determinations of $\vus$ from leptonic $K\mu2$ and semileptonic $Kl3$ kaon decays.
In the end a $3\,\sigma$ deficit in the CKM unitarity
 relation corresponding to the first row of this matrix (and less significant for  the first column) exists:
 \be\label{column}
 \vud^2+\vus^2+\vub^2=0.9985(5),\qquad \vud^2+\vcd^2+\vtd^2=0.9970(18)\,.
 \ee
 As reviewed in \cite{Crivellin:2022ctt} numerous authors made efforts
 in the literature to find the origin of this anomaly and to remove it
 with the help of NP models containing $W^\prime$, vector-like
 leptons, vector-like quarks, scalars, $Z^\prime$ gauge bosond and leptoquarks.

 Here we want to emphasize the following points:
 \begin{itemize}
 \item
   In the {\em absence} of new quarks, that must be vector-like, CKM unitarity
   {\em cannot} be violated. The violation in this case is only apparent due to possible contributions of bosons like $Z^\prime$ to decays used to determine
   $\vud$ and $\vus$ or due to hadronic uncertainties or wrong measurements.
   Otherwise GIM mechanism would fail and moreover at one-loop level, as illustrated below,  gauge dependences would be present. This also applies to the suppression mechanism presented here.
 \item
   In the {\em presence} of vector-like quarks CKM unitarity {\em can} be violated with the CKM matrix being submatrix of a unitary matrix involving SM quarks and vector-like quarks. Adding then diagrams with quark and vector-like quark
   exchanges the GIM mechanism and the suppression mechanism presented here
   are recovered.
 \end{itemize}

 Let us illustrate this problematic repeating the 1974 Gaillard-Lee
 calculation \cite{Gaillard:1974hs} of the  $K_L-K_S$ mass difference in the SM
 that was performed in the Feynman gauge.

The basic formula for the relevant Hamiltonian used to calculate $\Delta M_K$ in the SM 
that allows to reach our goal can be found in equation (6.67) in \cite{Buras:2020xsm}.
It reads
\be\label{Htot1}
\boxed{\mathcal{H}^{ij}_{\rm eff}=\frac{G_F^2}{16\pi^2}M_W^2 \lambda_i \lambda_j \left[(1+\frac{x_ix_j}{4})\,T(x_i,x_j) -2x_ix_j\tilde T(x_i,x_j)\right] (\bar s d)_{V-A}(\bar s d)_{V-A}\,,}  ~~~~~~~~~~~~~~~~~~~~~~~~~~~
\ee
with
\be
\lambda_i =V_{is}^*V_{id},\qquad  x_i=\frac{m_i^2}{M^2_W}, \qquad i=u,c,t,
\ee
where $V_{ij}$ are the elements of the CKM matrix.
The functions $T(x_i,x_j)$ and $\tilde T(x_i,x_j)$ are given as follows
\begin{align}\label{HWW}
T(x_i,x_j)&=& \left[\frac{x^2_i\log x_i}{(1-x_i)^2(x_i-x_j)}+
  \frac{x^2_j\log x_j}{(1-x_j)^2(x_j-x_i)}+\frac{1}{(1-x_i)(1-x_j)}\right]\,,
~~~~~~~~~~~~~~~~~~~~~~~~~~~~~~~~~~~
\\
T(x_i,x_i)&=&\left[\frac{2x_i\log x_i}{(1-x_i)^3}+
\frac{1+x_i}{(1-x_i)^2}\right]\,,~~~~~~~~~~~~~~~~~~~~~~~~~~~~~~~~~~~~~~~~~~~~~~~~~~~~~~~~~~~~~~~~~~~~~~~~~~~~~~~~~~~
\\
\tilde T(x_i,x_j)&=& \left[\frac{x_i\log x_i}{(1-x_i)^2(x_i-x_j)}+
\frac{x_j\log x_j}{(1-x_j)^2(x_j-x_i)}+\frac{1}{(1-x_i)(1-x_j)}\right]\,,~~~~~~~~~~~~~~~~~~~~~~~~~~~~~~~~~~~~\\
\tilde T(x_i,x_i)&=&\left[\frac{(1+x_i)\log x_i}{(1-x_i)^3}+
\frac{2}{(1-x_i)^2}\right]\,.~~~~~~~~~~~~~~~~~~~~~~~~~~~~~~~~~~~~~~~~~~~~~~~~~~~~~~~~~~~~~~~~~~~~~~~~~~~~~~
\end{align}
Detailed derivation of these formulae can be found in Section 6.2 of \cite{Buras:2020xsm}. 

Summing 
over the internal up-quarks and using the relation
\be\label{UTR}
\lambda_u+\lambda_c+\lambda_t=0\,,
\ee
that follows from the unitarity of the CKM matrix, we find the well known
SM expressions. For our purposes it is sufficient to keep only the term
involving $\lambda^2_c$ because this term dominates the SM contribution
to $\Delta M_K$. Using the standard formulae one finds then
\be\label{DMK1}
\Delta M_K=\frac{G_F^2}{6\pi^2}\hat B_K F_K^2M_W^2\eta_1 \lambda_c^2 x_c\,,
\ee
where the mass of the up-quark has been set to zero. $\eta_1$ is known at NNLO  in QCD and is subject to a large uncertainty but this
is not relevant for our main argument. Other factors are also well known.

On the other hand, not using (\ref{UTR}) and keeping only mass-independent terms  and $\ord(m^2_c/M_W^2)$ terms we find
\be\label{AJB}
\Delta M_K=2.52 \, T(x_c)\, 10^{-10}\GeV
\ee
where
\be\label{TXC}
T(x_c)= (\lambda_u+\lambda_c)^2+\lambda^2_c F_1(x_c)+2\lambda_u\lambda_c F_2(x_c)
\ee
with
\be
F_1(x_c)=2 x_c \log x_c + 3 x_c,\qquad F_2(x_c)= x_c \log x_c +  x_c.
\ee
To my knowledge these expressions are presented here for the first time.

Going back to 1974 and using the unitarity relation valid for
two quark generations
\be\label{UTR2}
\lambda_u+\lambda_c=0\,,
\ee
instead of (\ref{UTR}), we find 
$T(x_c)=\lambda_c^2 x_c $ and consequently (\ref{DMK1}).
However, if this relation is violated because of the CAA, the first mass independent contribution
and in particular the logarithmic terms, that do not cancel each other, have a large impact on the final result. Indeed
we find
\begin{align}\label{FF}
  F_1(x_c)&=&-41.4 \cdot 10^{-4} +7.5 \cdot 10^{-4}=-33.9\cdot10^{-4} \,,~~~~~~~~~~~~~~~~~~~~~~~~~~~~~~~~~~~~~~~~~~~~\\
  F_2(x_c)&=& -20.7 \cdot 10^{-4} +2.5\cdot 10^{-4}= -18.2\cdot10^{-4}\,.~~~~~~~~~~~~~~~~~~~~~~~~~~~~~~~~~~~~~~~~~~~~
\end{align}
We observe that the unitarity relation (\ref{UTR2}) is very important in
cancelling the logarithmic terms that are roughly by an order of magnitude
larger than than the SM result for $\Delta M_K$. But also  the modification due to the first term in (\ref{TXC}), that is not mass suppressed, is important.

However, these are minor problems in comparison with the fact that these results are gauge dependent and the ones just presented correspond to the Feynman gauge with the
gauge parameter $\xi=1$. Concentrating on the massless limit and repeating
the calculation in an arbitrary covariant gauge we find
\be
T(0)=  (\lambda_u+\lambda_c)^2\,\left[1-2\left(1+\frac{\xi}{1-\xi}\ln\xi\right)
  +\frac{1}{4}\left(1+\xi +\frac{2\xi}{1-\xi}\ln\xi\right)\right].
\ee
We note that for $\xi=1$ we reproduce (\ref{TXC}) for $x_c=0$ and that in the unitary gauge for which $\xi\to\infty$, $T(0)$ diverges. It vanishes only
if the unitarity relation in  (\ref{UTR2})is imposed, the crucial property
in the GIM mechanism.

Let us next return to the SM and consider three generations and the
$B^0_{s,d}-\bar B^0_{s,d}$ mass differences $\Delta M_{s,d}$. 
In this case the well known unitarity relations apply
\be\label{CKM}
\lambda_u+\lambda_c+\lambda_t= 0 \qquad \lambda_i=V^*_{ib}V_{iq},\qquad q=s,d\,.
\ee
If these relations are violated the results for $\Delta M_{s,d}$ are gauge dependent and divergent in the unitary gauge. In fact this has been demonstrated already in 2004 \cite{Buras:2004kq}  in the context  of the analysis in the littlest Higgs Model  in which CKM unitarity is violated due to the presence of a heavy
vector-like quark $T$ that mixes with the ordinary quarks.

In order to obtain gauge independent and finite results the CKM unitarity relation in (\ref{CKM}) should be replaced by the new one that reads
\be\label{GCKM}
\hat\lambda_u+\hat\lambda_c+\hat\lambda_t+\hat\lambda_T = 0 \qquad \hat\lambda_i=\hatV^*_{ib}\hat V_{iq}, \qquad q=s,d\,,
\ee
with $\hat V_{ij}$ containing corrections from the mixing with $T$ in question.
Explicit formulae for $\hat V_{ij}$ are given in \cite{Buras:2004kq}.
To my knowledge this was the first paper which explicitly demonstrated
 that only
 after the imposition of this unitarity relation
 the divergences in the unitary gauge could be removed.

 But now comes an important point made in \cite{Buras:2004kq}.
 In order for the relation in (\ref{GCKM}) to be effective it is crucial
 to include box diagrams involving both SM quarks and the vector-like quark $T$
 in addition to the usual box-diagrams involving ordinary quarks only
 and the ones that involve $T$ only. This is evident from the expressions above
 and applies to all VLQ models. More recent calculations of this type
 in the context of vector-like models can be found in
 \cite{Belfatto:2019swo,Belfatto:2021jhf,Botella:2021uxz,Crivellin:2022rhw}.

 \unboldmath
\noindent
\section{Summary}\label{LAST}
In this paper, we have introduced two
 $3\times3$ unitary and hermitian 
    matrices $\hat\Delta_d(Z^\prime)$ and $\hat\Delta_u(Z^\prime)$ that
    govern the flavour interactions mediated by $Z^\prime$ between   down-quarks and up-quarks,  respectively. This assures the suppression of these contributions to all  $Z^\prime$ mediated  FCNC processes at the one-loop level. 
    The explicit parametrization for $\hat\Delta_d(Z^\prime)$ is given in (\ref{B1}).  $\hat\Delta_u(Z^\prime)$ has the same structure but the mixing parameters
    and the two phases can differ from the ones in (\ref{B1}) if the couplings
    are right-handed. For left-handed couplings these matrices must be equal
    to each other, as given in (\ref{SU2}),
    due to  the unbroken ${\rm SU(2)}_L$ gauge symmetry in the SMEFT.
    Of interest  is also the breakdown of flavour universality in flavour conserving $Z^\prime$ couplings between the first two generations and the third one for both quarks and leptons. It is likely very small but this has to be investigated in the future.
    
    Due to the strong suppression of $Z^\prime$  contributions to all FCNC processes at the one-loop level, it is now fully justified to perform phenomenology
    at the tree level and determine the four parameters in (\ref{PARB}).
As according to \cite{Buras:2022qip}
       the CKM parameters can be determined from $\Delta F=2$
       processes alone without NP infection, one can begin the analysis
       with the resulting values of the  CKM parameters. They
read \cite{Buras:2022wpw}
\be\label{CKMoutput}
{\vcb=42.6(4)\times 10^{-3}, \quad 
\gamma=64.6(16)^\circ, \quad \beta=22.2(7)^\circ, \quad \vub=3.72(11)\times 10^{-3}\,.}
\ee
Having them, the four parameters in (\ref{PARB}) can be determined from
$\Delta F=1$ processes making sure that the infection of $\Delta F=2$ observables is within the uncertainties in (\ref{CKMoutput}). For the $K$ system, as
demonstrated in  \cite{Aebischer:2023mbz}, this can be achieved through
the choice $\delta_2-\delta_1=90^\circ$. But for the $B$-system it is more
challenging.

One   possibility would be the inclusion of right-handed couplings in addition to left-handed ones. This allows to  suppress NP contributions from  $Z^\prime$
  gauge boson to  $\Delta F=2$ observables but requires 
 fine tuning between left-left, right-right 
  and left-right operators contributing to these observables
  \cite{Buras:2014sba,Buras:2014zga,Crivellin:2015era}.
  Fortunately, in the case of $B$ physics this tuning is much smaller
  than in the case of $K$ physics, where this tuning is not required
  as demonstrated in \cite{Aebischer:2023mbz}.

  Much more efficient and more elegant solution to these problems is
  the introduction of a second $Z^\prime$ which collaborates with the
  first one to remove NP contributions from quark mixing and the
  two gauge bosons collaborate in the explanation of  the observed anomalies in $b\to s\mu^+\mu^-$ transitions.
  In this $Z^\prime$-Tandem scenario the new parameters in (\ref{PARB})
  are determined exclusively from $\Delta F=1$ processes with the CKM
  parameters determined already through $\Delta F=2$ processes and
  given in (\ref{CKMoutput}).  $Z^\prime$-Tandem scenario is proposed and
  discussed in \cite{Buras:2023xby}. The parametrization in (\ref{B1})
  presented here plays the crucial role in this new framework.

  The structure of the $\hat\Delta(Z^\prime)$
implies a number of correlations between $b\to s\mu^+\mu^-$, $B_{s,d}\to \ell^+\ell^-$, $\kpn$ and $\klpn$ branching ratios and other branching ratios as discussed in Section~\ref{Correlations}. Also correlations with the ratio
$\epe$ are present. They depend additionally on the 
leptonic couplings.With a sufficient number of observables, like the ones present in Flavio \cite{Straub:2018kue} and HEPfit  \cite{DeBlas:2019ehy} codes,
powerful tests of this scenario can be made and possible anomalies explained.
Even more important would be the construction of specific UV completions
which would allow to make more concrete predictions for FCNC observables
than in the simplified scenario presented here. In particular, the issue of
gauge anomalies related to the presence of new gauge bosons and addressed
in 331 models will likely require the introduction of new heavy fermions. But
as the cancelation of gauge anomalies is mass independent, we do not expect
significant modifications of this framework if their masses are sufficiently
high. Yet, this issue requires an additional study that is beyond the scope
of the present paper. 

Most important would be the discovery of $Z^\prime$
gauge boson at the LHC, but even if this will not happen, measurements of many observables will allow to test this scenario with explicit parametrizations of mixing matrices for quarks and leptons through virtual effects
alone. 
  
  \smallskip
  
  {\bf Acknowledgements}
  I would like to thank Herbert Spohn and Peter Weisz for  invaluable advices
  during the construction of the parametrization in (\ref{B1}). Discussions
  with Monika Blanke, Gerhard Buchalla and Andreas Crivellin are highly appreciated.
 Financial support from the Excellence Cluster ORIGINS,
funded by the Deutsche Forschungsgemeinschaft (DFG, German Research
Foundation), Excellence Strategy, EXC-2094, 390783311 is acknowledged.

\renewcommand{\refname}{R\lowercase{eferences}}

\addcontentsline{toc}{section}{References}

\bibliographystyle{JHEP}

\small

\bibliography{Bookallrefs}

\end{document}